\documentclass{PoS}

\def\lsc{L_8}

\title{All-order colour structure and two-loop anomalous dimension of 
soft radiation in heavy-particle pair production at the LHC}

\ShortTitle{Resummation of heavy-particle pair production at the LHC}

\author{\speaker{M.~Beneke}
        \thanks{Preprint numbers: TTK-10-11, SFB/CPP-10-13,
          IPPP/10/03, DCPT/10/06, FR-PHENO-2010-006. The work of M.B. is 
supported by the DFG Sonder\-forschungs\-be\-reich/Trans\-regio~9 
``Computergest\"utzte Theoretische Teilchenphysik''.}\\
        Institut f\"ur Theoretische Physik E, 
        RWTH Aachen University,\\  D - 52056 Aachen, Germany
        }

\author{P.~Falgari\\
        IPPP, Department of Physics, University of Durham, \\
        Durham DH1 3LE, England
        }

\author{C.~Schwinn\\
        Albert-Ludwigs Universit\"at Freiburg,\\
        Physikalisches Institut, D-79104 Freiburg, Germany
        }

\abstract{ 
We consider the factorization and resummation of soft and Coulomb gluons 
for pair-production processes of heavy coloured particles at hadron
colliders, and discuss: the construction of a colour basis that diagonalizes 
the leading soft function to all orders in perturbation theory; the 
determination of the two-loop soft anomalous dimension needed for NNLL 
resummations; and a general formula that provides all logarithmically 
enhanced ${\cal O}(\alpha_s^2)$ correction requiring as 
process-independent input only the one-loop hard matching coefficients.}

\FullConference{RADCOR 2009 - 9th International Symposium on 
Radiative Corrections
(Applications of Quantum Field Theory to Phenomenology) \\
                 October 25-30 2009\\
                 Ascona, Switzerland}

\begin{document}

\section{Introduction}

\noindent
Like in the well-studied Drell-Yan process, the 
partonic cross sections $q\bar q, q g,gg \to H H^\prime+X$ of 
heavy-particle pair production contain higher-order terms
\begin{equation}
\left[\alpha_s\ln^2\beta\right]^n,\qquad 
\beta^2=1-z=1-M_{H H^\prime}^2/\,\hat{s}
\end{equation}
in their perturbative expansion, 
which should be summed to all orders, if it can be 
argued that the hadronic cross section is dominated numerically by 
these threshold logarithms. For two coloured particles  in the 
final state with some fixed invariant mass $M_{H H^\prime}$
complications arise relative to the Drell-Yan process 
due to colour-exchange and kinematics-dependent anomalous 
dimensions similar to di-jet production~\cite{Kidonakis:1997gm}. 
For the total partonic cross section the only parametrically enhanced 
logarithms arise from the true production threshold, 
$M_{H H^\prime} = M_H+M_{H^\prime}$ and the kinematical dependence 
disappears. On the other hand, the particles are non-relativistic 
in this region and even more strongly enhanced terms 
$(\alpha_s/\beta)^n$ appear due to the Coulomb force. 
Although resummation for total partonic cross sections 
has been performed in the past~\cite{Bonciani:1998vc}, 
the issue of factorization of soft and Coulomb gluons, and their 
simultaneous resummation has not been addressed with rigour 
until recently. In this proceedings article we discuss the factorization 
formula applicable to this situation, the diagonal colour basis 
for the leading soft function relevant to the heavy-particle 
pair production process, and the two-loop anomalous dimension 
for soft radiation. For details 
we refer to~\cite{Beneke:2009rj}. See the talk by 
P.~Falgari~\cite{Beneke:2009nr} for a discussion of resummation 
of the squark anti-squark production cross section at the LHC.

To define the LL, NLL, etc. approximations of the resummed 
cross section in the presence of Coulomb effects, 
we note that near threshold 
the usual expansion, where $\alpha_s \ln\beta$ in the exponent 
of~(\ref{eq:syst}) below counts as order one, 
is combined with an expansion in $\beta$, such that $\alpha_s/\beta$ 
also counts as order one. This leads to a parametric representation 
of the expansion of the cross section in the form
\begin{eqnarray}
\label{eq:syst}
\hat{\sigma}_{p p'} &=& \,\hat \sigma^{(0)}\, 
\sum_{k=0} \left(\frac{\alpha_s}{\beta}\right)^k \,
\exp\Big[\underbrace{\ln\beta\,g_0(\alpha_s\ln\beta)}_{\mbox{(LL)}}+ 
\underbrace{g_1(\alpha_s\ln\beta)}_{\mbox{(NLL)}}+
\underbrace{\alpha_s g_2(\alpha_s\ln\beta)}_{\mbox{(NNLL)}}+\ldots\Big]
\nonumber\\[0.2cm]
&& \,\times
\left\{1\,\mbox{(LL,NLL)}; \alpha_s,\beta \,\mbox{(NNLL)}; 
\alpha_s^2,\alpha_s\beta,\beta^2 \,\mbox{(NNNLL)};
\ldots\right\}, 
\end{eqnarray}
which reproduces the standard structure~\cite{Bonciani:1998vc} away from
threshold for $k=0$ and no expansion in $\beta$. This implies that at 
${\cal O}(\alpha_s^2)$ relative to the Born cross section 
the terms $\alpha_s^2\times\{1/\beta; \ln^{2,1} \beta ; 
\beta \times \ln^{4,3} \beta \}$ are NNLL. Note that   
$\alpha_s^2 \ln^{2,1}\beta$ terms may arise from the product of a 
Coulomb-en\-han\-ced one-loop correction $\alpha_s/\beta$ and 
a $\beta$-suppressed soft emission $\alpha_s\beta\ln^{2,1}\beta$. 
This highlights the subtle point that one must consider approximations 
one order beyond the standard eikonal approximation to capture all 
NNLL terms. 

\section{Factorization of soft and Coulomb gluons}

\noindent
The factorization of soft and Coulomb gluons is a non-trivial issue, since 
the non-relativistic energy of the heavy particles is of the same order 
as the soft gluon momentum. Hence, Coulomb exchanges are not part of 
the hard process and take place (diagrammatically) ``in between''  the 
soft gluon emissions. In~\cite{Beneke:2009rj} it is shown that both 
effects are simultaneously resummed by means of the formula
\begin{equation}
\hat\sigma(\beta,\mu)
= \sum_a \sum_{i,i'}H_{ii'}^a(M,\mu)
\;\int d \omega\;
\sum_{R_\alpha}\,J_{R_\alpha}^a(E-\frac{\omega}{2})\,
W^{a,R_\alpha}_{ii'}(\omega,\mu),
\label{factform}
\end{equation}
which contains a multiplicative short-distance coefficient 
$H_{ii'}^a$ in each colour (in higher orders also spin) configuration, 
and a convolution of soft functions $W^{a,R_\alpha}_{ii'}$ with 
Coulomb functions $J_{R_\alpha}^a$. The factorization of soft gluons 
from collinear fields follows from a field redefinition with a light-like 
Wilson line in soft-collinear effective theory (SCET) in the usual 
way~\cite{Bauer:2001yt}. To prove the decoupling from non-relativistic 
fields we redefine $\psi_a(x)=S_v^{(R)}(x^0)_{ab}\,\psi_b^{(0)}(x)$ 
with a time-like Wilson line  
\begin{equation}
S^{(R)}_{v}(x) =
\overline{\mbox{P}}\exp \left[-i g_s \int_0^{\infty} d t \; v\cdot A^c_s(x+v t)
  {\bf T}^{(R)c} \right].
\end{equation}
This has the effect of turning $D_s^0$ into $\partial^0$ in the 
leading-order PNRQCD Lagrangian
\begin{eqnarray}
\label{eq:NRQCD}
\mathcal{L}_{\mbox{\tiny PNRQCD}} &=&
\psi^\dagger \left(i D_s^0+\frac{\vec{\partial}^2}{2 m_H}
 +\frac{i\Gamma_{H}}{2}\right) \psi
+\psi^{\prime\,\dagger} \left(i D_s^0+\frac{\vec{\partial}^2}{2 m_{H'}}
 +\frac{i\Gamma_{H^\prime}}{2}\right) \psi'
\nonumber \\
&&+\,\int d^3 \vec{r}\, 
\left[\psi^{\dagger}{\bf T}^{(R)a} \psi\right]\!(\vec r\,)
\left(\frac{\alpha_s}{r}\right)
\left [\psi^{\prime\,\dagger}{\bf T}^{(R')a}\psi^\prime\right]\!(0).
\end{eqnarray}
The key observation is that $S_v$ drops out from the Coulomb interaction 
expressed in terms of the new fields, since 
$S_v^{(R)\dagger} {\bf T}^{(R)a}S_v^{(R)} = [S_{\rm ad}]^{ab}
\,{\bf T}^{(R)b}$ in any representation $R$ and since $S_{\rm ad}$ in the 
adjoint representation is real and independent of $\vec{r}$. For the 
latter point it is important that the soft gluon field in $D_s^0$ depends 
only on $x^0$~\cite{Beneke:1999zr}, while the Coulomb interaction is 
non-local but instantaneous. This proves decoupling of soft gluon and 
Coulomb resummation, since soft 
gluons disappear from the leading-order Lagrangians for the other 
fields. They do not disappear from higher-order terms in the 
SCET and PNRQCD Lagrangians, but these sub-leading interactions can be 
treated as perturbations in $\beta$, resulting in the sum over $a$ 
in (\ref{factform}).

\section{All-order diagonal colour basis 
for the leading soft function}

\noindent
The soft function relevant to the leading power in the 
$\beta$ expansion ($a=1$ in (\ref{factform})) 
is given by the Fourier transform of 
$\hat W^{R_\alpha}_{ii'}(z,\mu)=  P^{R_\alpha}_{\{k\}}
c^{(i)}_{\{a\}} \hat W^{\{k\}}_{\{ab\}}(z,\mu)c^{(i')*}_{\{b\}}$ 
where 
\begin{equation}
\hat W^{\{k\}}_{\{ab\}}(z,\mu)
=\langle 0|\overline{\mbox{T}}[ 
S_{v,b_4 k_2} S_{v,b_3 k_1} S^\dagger_{\bar{n},jb_2} S^\dagger_{n, ib_1}](z)
\mbox{T}[S_{n,a_1i} S_{\bar{n},a_2j} S^\dagger_{v,k_3 a_3} 
S^\dagger_{v,k_4 a_4}](0)|0\rangle,
\label{ww}
\end{equation}
and $\{a\}=a_1 a_2 a_3 a_4$ denotes a colour multi-index for 
the $2\to 2$ production process. $\hat W^{R_\alpha}_{ii'}(z,\mu)$ 
is obtained from (\ref{ww}) by projecting it with
$P^{R_\alpha}_{\{k\}}$ on 
an irreducible representation $R_\alpha$ in the product 
$R\otimes R'=\sum_{\beta} R_{\beta}$ of final state representations, 
and on the elements $c^{(i)}_{\{a\}}$ of a suitable colour basis. 
The number of basis elements ($i, i^\prime =1 \ldots n$) 
is constrained by colour conservation. 
Further decomposing the product of initial state colour 
representations into irreducible ones according to 
$r\otimes r' =\sum_{\alpha} r_\alpha$, the number of basis elements 
equals the number of pairs 
$P_i=(r_{\alpha}, R_{\beta})$ of equivalent 
representations $r_\alpha$ and $R_\beta$. For example, in case of a  
$8\otimes 8 \to 8\otimes 8$ process, 
$P_i\in 
\{(1,1),\; (8_S,8_S),\;(8_A,8_S),\;(8_A,8_A),$ $\;(8_S,8_A),\;\;(10,10),\;
(\overline{10},\overline{10}),\; (27,27) \}$, so the basis is 
eight-dimensional. One then finds~\cite{Beneke:2009rj} that 
$W^{R_\alpha}_{ii'}(\omega,\mu)$ in (\ref{factform}) is diagonal 
to all orders in perturbation theory in the colour basis
\begin{equation}
c_{\{a\}}^{(i)}=\frac{1}{\sqrt{\mbox{dim}(r_\alpha)}}\,
  C^{r_\alpha}_{\alpha a_1a_2} C^{R_{\beta}\ast}_{\alpha a_3a_4}
\end{equation}
constructed from the Clebsch-Gordan 
coefficients that couple the initial and final colour representations 
to the equivalent pairs $P_i$. This result follows from colour 
conservation; the fact that the Coulomb interaction is diagonal 
in the irreducible colour representations; and the ability to 
combine the two final-state Wilson lines into a single one using
$C^{R_\alpha}_{\alpha a_1a_2}S^{(R)}_{v,a_1b_1}
S^{(R')}_{v,a_2b_2}=S^{(R_\alpha)}_{v,\alpha\beta}
C^{R_\alpha}_{\beta,b_1b_2}$, since the Wilson lines for a heavy-particle 
pair produced directly at threshold carry the same velocity vector $v$. 
Finally,  due to Bose symmetry of the soft function there is no
interference of the production from a symmetric and antisymmetric 
colour octet, which excludes the only possible off-diagonal terms 
after applying the arguments above.

\section{Two-loop anomalous dimension 
for heavy-particle pair production}

\noindent 
Once the leading soft function is diagonal and reduced to the soft function 
of an effective $2\to 1$ process with a coloured final-state 
particle in representation $R_\alpha$, the soft gluon part of the 
resummation can be done in SCET just as for Drell-Yan 
production~\cite{Becher:2007ty}. For NNLL resummation, one only 
needs in addition the two-loop anomalous dimension and the one-loop 
finite term of the 
heavy-particle soft functions. 
Since (\ref{ww}) is essentially the ``square'' of 
the soft function relevant to the effective $2 \to 1$ amplitude, the 
two-loop anomalous dimension satisfies Casimir scaling and 
can be extracted from \cite{Becher:2009kw}. In particular, potential 
three-particle colour correlations vanish trivially after the 
reduction to the $2\to 1$ process. The result can be 
converted to the anomalous dimension required in the Mellin-space 
resummation formalism~\cite{Bonciani:1998vc}, which requires the one-loop 
calculation of the soft function, and one 
finds~\cite{Beneke:2009rj} 
\begin{equation}
 D_{HH'}^{(1)R_\alpha}=-C_{R_\alpha}C_A\left(\frac{460}{9}-
\frac{4\pi^2}{3}+8\zeta_3\right)
 +\frac{176}{9}C_{R_\alpha} T_F n_f,
\end{equation}
which has been confirmed independently in~\cite{Czakon:2009zw}.
The process-independent ingredients for NNLL resummation of 
threshold logarithms in arbitrary production processes of heavy 
coloured particles at hadron colliders are now in place. 
The resummation of the squark anti-squark production cross section at the LHC 
at NLL using the formalism outlined above is discussed 
in~\cite{Beneke:2009nr}.

\section{Threshold enhancements at 
NNLO}

\noindent
When resummation is not required one can use the fixed-order expansion 
of (\ref{factform}) to calculate the velocity-enhanced terms at NNLO. 
However, to obtain all $\alpha_s^2 \ln^{2,1}\beta$ correction terms, one must 
include logarithms from sub-leading heavy-quark potentials and sub-leading 
soft gluon couplings~\cite{Beneke:2009rj}. These additional terms have 
been worked out explicitly in~\cite{Beneke:2009ye}. The contributions 
from sub-leading soft gluon couplings could be a source of 
velocity-enhanced three-particle 
colour correlations, which have been calculated in~\cite{Ferroglia:2009ii} 
and found to be non-zero at the amplitude level. 
The result of~\cite{Beneke:2009ye} implies that there are no 
contributions to the 
$\ln\beta$ terms from such three-particle correlations in both, the 
virtual and real corrections to the total cross section. This holds 
independent of particular colour representations.

Here we provide the velocity-enhanced terms at NNLO for the 
production of a pair of heavy particles with equal mass $m$ 
in the scattering 
of massless partons in colour representations $r$ and $r'$, 
respectively, under the only assumption that the Born cross section 
admits an $S$-wave term proportional to $\beta$.
The heavy-particle pair is in colour representation 
$R_\alpha$ and a definite spin state (singlet or triplet for 
spin-1/2 fermions). Denoting by $\sigma^{(2)}_X$ the NNLO correction 
relative to the Born cross section, the threshold expansion 
reads~\cite{Beneke:2009ye} 
\begin{eqnarray}
\sigma^{(2)}_X &\!\!=\!\!& 
\frac{4 \pi^4 D_{R_\alpha}^2}{3 \beta^2} +
\frac{\pi^2 D_{R_\alpha}}{\beta} \,
\bigg\{(-8) \,(C_r+C_{r'}) \bigg[\ln^2\left(\frac{2 m\beta^2}{\mu}\right)
-\frac{\pi^2}{8}\bigg]
+ 2 \,(\beta_0+4  C_{R_\alpha}) \ln\left(\frac{2 m\beta^2}{\mu}\right)
\nonumber\\
&& \hspace*{1cm} 
- \,8  C_{R_\alpha}-2 a_1 -4 \,\mbox{Re}\,[C^{(1)}_X] 
+2 \beta_0  \ln\left(\frac{2 m}{\mu}\right)
\bigg\}
\nonumber\\
&& 
+\, 128 \,(C_r+C_{r'})^2 \ln^4 \beta 
+ 64 \,(C_r+C_{r'}) \left\{4 \,(C_r+C_{r'}) 
\left(\lsc-2\right) -\frac{\beta_0}{3} -2 C_{R_\alpha}\right\} 
\ln^3\beta
\nonumber\\
&&+\,\bigg\{ \frac{8}{3} \,(C_r+C_{r'})^2 
\left[72  \lsc^2   -288  \lsc + 576-35 \pi^2\right]
+\frac{16}{9} \,(C_r+C_{r'}) \,\Big[18 \,{\rm Re}\,[C^{(1)}_X] 
\nonumber\\
&& \hspace*{0.6cm} 
+\, 18 \beta_0 \left(-\lsc+2\right) 
+36 C_{R_\alpha} \left(-3\lsc+7\right)+C_A (67-3 \pi^2)-20 n_l T_f
\Big]
\nonumber\\
&&\hspace*{0.6cm} 
+\,16 C_{R_\alpha} (\beta_0+2 C_{R_\alpha})
\phantom{\hspace{-0.2cm}\frac{|}{|}}\bigg\} \ln^2 \beta 
\nonumber\\
&&
+ \,\bigg\{ 8 \,(C_r+C_{r'})^2 \left[
8 \lsc^3 -48\lsc^2 +\left(192-\frac{35 \pi^2}{3}\right)
\lsc -384+\frac{70 \pi^2}{3}+112 \zeta_3 \right] 
\nonumber\\
&&\hspace*{0.6cm}
+\,2 \,(C_r+C_{r'}) \left[-16 \,\mbox{Re}\,[C^{(1)}_X] \left(-\lsc+2\right) 
+\beta_0 \left(-8\lsc^2+32 \lsc -64+\frac{11 \pi^2}{3}\right)\right.  
\nonumber\\
&&\hspace*{1cm} 
+\,2 C_{R_\alpha} \left(-24 \lsc^2+112 \lsc -224 
+\frac{35\pi^2}{3}\right)
\nonumber\\
&&\hspace*{1cm}  
+ \, C_A \left(\frac{8}{3} \left(\frac{67}{3}-\pi^2\right)\lsc
  -\frac{4024}{27}+\frac{59 \pi^2}{9}+28 \zeta_3\right)
\left.  +\frac{4 n_l T_f}{9} \left(\!-40\lsc+\frac{296}{3}-\pi^2\right)\right]
\nonumber\\
&&\hspace*{0.6cm}
+\,4 \,C_{R_\alpha} 
\bigg[ -4 \,\mbox{Re}\,[C^{(1)}_X] 
-4 \,(\beta_0+2 C_{R_\alpha}) \left(-\lsc+3\right)
+C_A \left(-\frac{98}{9}+\frac{2 \pi^2}{3}-4 \zeta_3
    \right)
\nonumber\\
&&\hspace*{1cm}
+\frac{40}{9} n_l T_f \bigg]
+16 \pi^2 D_{R_\alpha} \Big[C_A-2 D_{R_\alpha} (1+v_{\rm spin}) \Big] 
\bigg\} \,\ln \beta+{\cal O}(1)\,. 
\label{eq:general}
\end{eqnarray}
Here $C_r$, $C_{r'}$ and $C_{R_\alpha}$ denote the quadratic Casimir
operators of the colour representations, $\beta_0 =
\frac{11}{3} C_A-\frac{4}{3} n_l T_f$ is the one-loop beta-function 
coefficient, and $L_8 = \ln(8 m/\mu)$. The quantities $D_{R_\alpha}$, 
$a_1 =\frac{31}{9} C_A-\frac{20}{9} n_l T_f$ and $v_{\rm spin}$ 
are connected with the heavy-quark potentials such that 
$v_{\rm spin} = 0$ and $-2/3$ for a pair of spin-1/2 fermions in a 
spin-singlet and spin-triplet state, respectively, and 
$D_{R_\alpha}$ refers to the strength of the Coulomb potential in 
representation $R_\alpha$ ($D_{R_\alpha} = -C_F$ 
for the singlet and $D_{R_\alpha} = -(C_F-C_A/2)$ for the 
octet representation). The only process-specific input is 
$\mbox{Re}\,[C^{(1)}_X]$, equal to one half of the the one-loop 
short-distance coefficient $H_X(M,\mu)$ in (\ref{factform}), when the 
heavy-particle pair is in colour and spin state $X$. Alternative to a 
direct computation, 
it can also be deduced from the constant term in the 
threshold limit of the NLO 
production cross section $\sigma^{(1)}_X$ 
in colour and spin channel $X$ by comparing the 
expansion  of $\sigma^{(1)}_X$ to the formula 
\begin{eqnarray}
\sigma^{(1)}_X &\!\!=\!\!& 
- \frac{2 \pi^2 D_{R_\alpha}}{\beta} +
4\,(C_r+C_{r'}) \bigg[\ln^2\left(\frac{8 m\beta^2}{\mu}\right)
+8 -\frac{11 \pi^2}{24}\bigg]
\nonumber\\
&&  
- \,4 \,(C_{R_\alpha}+ 4\,(C_r+C_{r'})) \,
\ln\left(\frac{8 m\beta^2}{\mu}\right) 
+ 12  C_{R_\alpha} +2 \,\mbox{Re}\,[C^{(1)}_X] + 
{\cal O}(\beta)\,.
\label{eq:NLOexp}
\end{eqnarray}

\end{document}